# Memory-bit selective recording in vortex-core cross-point architecture

Young-Sang Yu,[1] Hyunsung Jung,[1] Ki-Suk Lee,[1] Peter Fischer,[2] and Sang-Koog Kim[1,a)]

[1]*Research Center for Spin Dynamics & Spin-Wave Devices, and Nanospinics Laboratory, Research Institute of Advanced Materials, Department of Materials Science and Engineering, Seoul National University, Seoul 151-744, South Korea*

[2]*Center for X-ray Optics, Lawrence Berkeley National Laboratory, Berkeley CA 94720, USA*

In our earlier work [Appl. Phys. Lett. **92**, 022509 (2008)], we proposed nonvolatile vortex random access memory (VRAM) based on the energetically stable twofold ground state of vortex-core magnetizations as information carrier. Here we experimentally demonstrate reliable memory bit selection and low-power-consumption recording in a two-by-two vortex-state dot array. The bit selection and core switching is made by flowing currents along two orthogonal addressing electrode lines chosen among the other crossed electrodes. Tailored pulse-type rotating magnetic fields are used for efficiently switching a vortex core only at the intersection of the two orthogonal electrodes. This robust mechanism provides reliable bit selection and information writing operations in a potential VRAM device.

[a)] The author to whom all correspondence should be addressed; Eelectronic mail: sangkoog@snu.ac.kr

An energy-efficient, ultrahigh-density, ultrafast, and nonvolatile solid-state universal memory is a long-held dream in information storage technologies. Magnetic vortices in patterned magnetic dots are a promising candidate for practical applications in storage devices[1], not only because of the energetically stable twofold ground state of their core magnetization[2-8] but also due to a well controllable low-power driven core switching.[9-17] However, technologically important information writing in an array of memory bits has not been demonstrated yet.

In this letter, towards realizing magnetic vortex random access memory (VRAM) conceptually proposed earlier,[9] we experimentally demonstrate reliable memory bit selection and its low-power-consumption recording using circular rotating fields of pulse type in the existing basic cross-point architecture.

In earlier works,[9,15,17] it was reported that circular-rotating magnetic fields are useful for efficient manipulation of the binary states of vortex-core orientation. This special magnetic field represents a magnetic field that rotates either counter-clockwise (CCW) or clockwise (CW) on the film plane with a single-harmonic angular frequency $\omega_H$, as illustrated in Fig.1. The ideal CCW and CW circular-rotating fields can be expressed as the vector sum of two harmonic oscillating fields. Such ideal circular fields are known to be most efficient for selectively switching either the upward or downward core with the lowest strength when the field frequency is tuned to the vortex angular eigenfrequency $\omega_D$. The CCW (CW) rotating field of $\omega_{\mathbf{H}} = \omega_D$ can switch only the upward (downward) core to its reversed orientation, which field is called resonant-mode rotating field.[18]

From the real application point of view, simple pulse currents such as those of the Gaussian type are much more useful than ideal circular-rotating fields, so that as a further step towards realizing VRAM,[9,19] we demonstrated memory bit selection and recording in a two-by-two dot array, for example, with the existing cross-point architecture scheme, as shown in Fig. 2(a). The sample used is a two-by-two Permalloy (Py: $Ni_{80}Fe_{20}$) disk array as shown in the microscope image [Fig. 2(b)]. The Py disks of radius $R = 2.5$ $\mu$m and thickness $L = 70$ nm were deposited by magnetron sputtering under base pressures of less than $5 \times 10^{-9}$ Torr, and then patterned by typical e-beam lithography (Jeol, JBX9300FS) and subsequent lift-off processes. Each disk was placed at the intersection of two Au electrodes and capped with 2-nm-thick Pd layers (without breaking the vacuum) to prevent oxidation. The four interaction areas of the crossed-stripline electrodes are 50 nm thick and 10 $\mu$m wide. In order to obtain sufficient soft x-ray transmission through the sample, the electrodes were deposited onto the 200-nm-thick silicon nitride membranes of a 5 mm-by-5 mm window by electron-beam evaporation under base pressures of less than $1\times10^{-8}$ Torr.

When using orthogonal and uinpolar Gaussian-pulse currents [see Fig. 2(c)] to switch vortex cores, it is necessary to optimize the pulse parameters of width $\sigma$ and time delay $\Delta t$ to obtain the lowest field strength for achieving low-power-consumption vortex-core switching. Through our earlier work,[20] the optimal values were found to to be $\sigma = 1/\omega_D$ and $\Delta t = \frac{\pi}{2} p/\omega_D$ where $p$ is the polarization of the initial vortex state, $p = + 1$ (- 1), corresponding to the upward (downward) core magnetization.[20]  Note that the optimal pulse parameters are determined only by $\omega_D$ for a given polarization $p$. Then, a mechanism for memory bit selection and switching based on the above mentioned orthogonal, Gaussian pulses with the optimal pulse parameters is to use the fact that the threshold field strengths $H_{th}$ for the different polarizations of pulse-type rotating fields are contrasting for a given upward-to-downward core switching. Figure 3 shows the analytical calculation for the same dimensions as those of the real sample. The purple (orange) color indicates the results with $\Delta t = +\frac{1}{2}\pi/\omega_D$  ($\Delta t = -\frac{1}{2}\pi/\omega_D$), whereas the green is the result with a single oscillating Gaussian pulse. The positive and negative values of $\Delta t$ correspond to a CCW and CW rotating field, respectively, while $\Delta t = 0$ corresponds to the linear field. As is apparent, the threshold fields varied markedly for the different field polarizations. These analytical calculations were obtained using $\omega_D / 2\pi = 146$ MHz for the sample dimensions. The $H_{\text{th}}$ of $\Delta t = +\frac{1}{2}\pi/\omega_D$ (CCW rotating field) is lowest at $\sigma =$ 1.09 ns for the upward-to-downward core switching. The mimimum $H_{\text{th}}$ for the linear field is higher by a factor of almost two than that for the CCW rotating field: 22.1 versus 11.3 Oe. The upward-to-downward core switching by the CW rotating field ( $\Delta t = -\frac{1}{2}\pi/\omega_D$ ) requires higher field strengths than 43 Oe at $\sigma = 1.09$ ns, which value is much higher than those of the CCW rotating and linear field strengths. This large difference enables reliable selection of a memory bit, without the typical half-selection problem, simply by choosing two electrodes at the intersection of which there is to be a vortex core in the basic cross-point architecture, as shown in Fig. 2(a).

Next, in order to experimentally demonstrate the above concept, we monitored the out-of-plane magnetizations of vortex core orientations, after or before their switching , by using high-resolution magnetic transmission soft x-ray microscopy (MTXM) through the x-ray magnetic circular dichroism (XMCD) contrast at the Fe $L_3$ edge.[21] The MTXM images around the core regions of vortices at locations marked as "b", "c", "d" are shown in Fig. 4. The dark and white spots indicate the upward and downward core orientations, respectively, in our experimental setup. By monitoring the dark and white spots in these experimental XMCD images, we could readily determine whether vortex-core switching events had occurred for a given field pulse.

Reliable selection of a memory bit was achieved by choosing the bit- and word-line electrodes, as schematically illustrated in Fig. 2(a). In the real sample, the minimum values of $H_{th}$ were obtained to be 11.0 Oe for case of using $\Delta t = +2.0$ (- 2.0) ns for the upward (downward) core switching.  By applying two pulse currents of the optimal values $\Delta t = +2.0$ ns (CCW rotating field) and $\sigma = 1.27$ ns [Ref. 22] along the two striplines marked as "$W_2$" and "$B_2$", only the upward core located at position "c" is switched. In this case, as the x-ray microscopy images provided make clear, only at the cross-point did the upward core orientation switch, once, with a further increase to 11.5 Oe, whereas the reversed downward core, according to the same CCW rotating field, did not switch even with further increases of $H_0$. Notably, neither the upward nor the downward core orientation at locations "b" and "c", switched with $H_0 = 11.5$ Oe, because this field strength of the linear fields applied at those locations was not sufficient to affect the switching of either the upward or downward core orientation. As already mentioned, the threshold field strength of the linear field is almost two times larger than that of the resonant-mode rotating field. This large difference in threshold strength between the resonant-mode rotating field and the linear field is promising for avoidance of the half-selection switching problem[23] occurring in the conventional MRAM structure.

The present experimental demonstration of low-power-consumption reliable recording of a selected memory bit in the cross-point architecture is promising for non-volatile information storage and writing operations. Thus, this work imparts further momentum to the realization of VRAM based on the unique vortex structure and its novel dynamic properties. This technological achievement based upon fundamentals of vortex-core switching dynamics is the critical milestone toward the realization of a VRAM device.

**Acknowledgements**

We are thankful to M.Y. Im for assistance at the beamline. This research was supported by the Basic Science Research Program through the National Research Foundation of Korea (NRF) funded by the Ministry of Education, Science, and Technology (Grant No. 20100000706). S.-K.K. was supported by the LG YONAM foundation under the Professors' Overseas Research Program. Use of the soft X-ray microscope was supported by the Director, Office of Science, Office of Basic Energy Sciences, Materials

Sciences and Engineering Division, U.S. Department of Energy.

## References

[1] R. P. Cowburn, Nature Mater. **6**, 255 (2007).

[2] A. Hubert and R. Schäfer, *Magnetic Domains: The analysis of Magnetic Microstructure* (Springer-Verlag, Berlin, 1998).

[3] J.Miltat and A. Thiaville, Science **298**, 555 (2002).

[4] J. Raabe, R. Pulwey, R. Sattler, T. Schweinböck, J. Zweck, and D. Weiss, J. Appl. Phys. **88**, 4437 (2000).

[5] T. Shinjo, T. Okuno, R. Hassdorf, K. Shigeto, and T. Ono, Science **289**, 930 (2000).

[6] A. Wachowiak, J. Wiebe, M. Bode, O. Pietzsch, M. Morgenstern, and R. Wiesendanger, Science **298**, 577 (2002).

[7] S.-K. Kim, J. B. Kortright, and S.-C. Shin, Appl. Phys. Lett. **78**, 2742 (2001).

[8] K. L. Metlov and K. Y. Guslienko, J. Magn. Magn. Mater. **242**, 1015 (2002).

[9] S.-K. Kim, K.-S. Lee, Y.-S. Yu, and Y.-S. Choi, Appl. Phys. Lett. **92**, 022509 (2008).

[10] B. Van Waeyenberge, A. Puzic, H. Stoll, K. W. Chou, T. Tyliszczak, R. Hertel, M. Fähnle, H. Brückl, K. Rott, G. Reiss, I. Neudecker, D. Weiss, C. H. Back, and G. Schütz, Nature (London) **444**, 461 (2006).

[11]K. Yamada, S. Kasai, Y. Nakatani, K. Kobayashi, H. Kohno, A. Thiaville, and T. Ono, Nature Mater. **6**, 269 (2007).

[12]R. Hertel, S. Gliga, M. Fähnle, and C. M. Schneider, Phys. Rev. Lett. **98**, 117201 (2007).

[13]K.-S. Lee, K. Y. Guslienko, J.-Y. Lee, and S.-K. Kim, Phys. Rev. B **76**, 174410 (2007).

[14]S.-K. Kim, Y.-S. Choi, K.-S. Lee, K. Y. Guslienko, and D.-E. Jeong, Appl. Phys. Lett. **91**, 082506 (2007).

[15]K.-S Lee, S.-K. Kim, Y.-S. Yu. Y.-S. Choi, K. Y. Guslienko, H. Jung, and P. Fischer, Phys. Rev. Lett. **101**, 267206 (2008).

[16]M. Weigand, B. Van Waeyenberge, A. Vansteenkiste, M. Curcic, V. Sackmann, H. Stoll, T. Tyliszczak, K. Kaznatcheev, D. Bertwistle, G. Woltersdorf, C. H. Back, and G. Schütz, Phys. Rev. Lett. **102**, 077201 (2009).

[17]M. Curcic, B. Van Waeyenberge, A. Vansteenkiste, M. Weigand, V. Sackmann, H. Stoll, M. Fähnle, T. Tyliszczak, G. Woltersdorf, C. H. Back, and G. Schütz, Phys. Rev. Lett. **101**, 197204 (2008).

[18]K.-S. Lee and S.-K. Kim, Phys. Rev. B **78**, 014405 (2008).

[19]S.-K. Kim, K.-S. Lee, Y.-S. Choi, and Y.-S. Yu, IEEE Trans. Mag. **44**, 3071-3074 (2008).

[20]Y.-S. Yu, K.-S. Lee, H. Jung, Y.-S. Choi, M.-W. Yoo, D.-S. Han, M.-Y. Im, P. Fischer, and S.-K. Kim, Phys. Rev. Lett., Submitted (2010).

[21]P. Fischer, D.-H. Kim, W. Chao, J. A. Liddle, E. H. Anderson, and D. T. Attwood, Mater. Today **9**, 26 (2006).

[22]This value experimentally used is different from that of the analytical calculation, $\sigma$ = 1.09 ns. The analytically derived $\sigma$ value was close to the experimentally optimized value, $\sigma$ = 1.27 ns. This seemingly is associated with the difference in the eigenfrequency between the real sample and the value of $\omega_D / 2\pi = 146$ MHz used for the analytical calculation.

[23]C. Chappert, A. Fert, and F. N. Van Dau, Nature Mater. **6**, 813 (2007).

**Figure captions**

FIG. 1. (Color online) Two crossed electrodes and Oersted fields induced by current flows along the two electrodes. Time-varying oscillating magnetic fields produced by two orthogonal alternating currents. Schematic illustration of ideal CCW and CW circular-rotating magnetic fields with a single harmonic angular frequency $\omega_\mathbf{H}$, produced locally at the intersection of the two electrodes., which are mathematically expressed as

$$\mathbf{H}_{\text{CCW}} = H_0 \cos(\omega_\mathbf{H} t)\hat{\mathbf{x}} + H_0 \sin(\omega_\mathbf{H} t)\hat{\mathbf{y}} \qquad \text{and} \qquad \mathbf{H}_{\text{CW}} = H_0 \cos(\omega_\mathbf{H} t)\hat{\mathbf{x}} - H_0 \sin(\omega_\mathbf{H} t)\hat{\mathbf{y}} \qquad,$$

where $\bar{H} = H_0$.

FIG. 2. (Color online) (a) Schematic illustration of two-by-two vortex-state Py disk array. (b) The optical microscopy image of the sample: each Py dot has $R = 2.5$ $\mu$m and $L = 70$ nm, and the Au crossed striplines are 10 $\mu$m wide at the interactions and 50 nm thick. (c) The parameters of the two orthogonal Gaussian pulse currents, width $\sigma$ and time delay $\Delta t$.

FIG. 3. (Color online) Analytical calculation of the boundary diagrams of vortex-core switching events on the plane of $\sigma$ and $H_0$ for initial up-core vortex state ($p = +1$) in a Py disk of the same dimensions as the real sample, $R = 2.5$ $\mu$m and $L = 70$ nm. Purple (orange) color indicates the results with $\Delta t = +\frac{1}{2}\pi/\omega_D$ ($\Delta t = -\frac{1}{2}\pi/\omega_D$), whereas the green is the result with a single oscillating Gaussian pulse, $I_{x0}$ ($I_{y0} =0$) or $I_{y0}$ ($I_{x0} =0$). The gray-colored vertical line is placed at $\sigma$ = 1.09 ns.

Fig. 4. (Color online) Zoomed images at the vortex core region in three different disks marked as "b", "c", and "d" in Fig 2, before or upon vortex-core switching by indicated types of applied fields. The field strength along the $x$ and the $y$ direction generated by each of the Gaussian pulses was $H_0 = 11.5$ Oe. The initial core magnetization was upward for all of the disks.

**Figures**

**FIG. 1.**

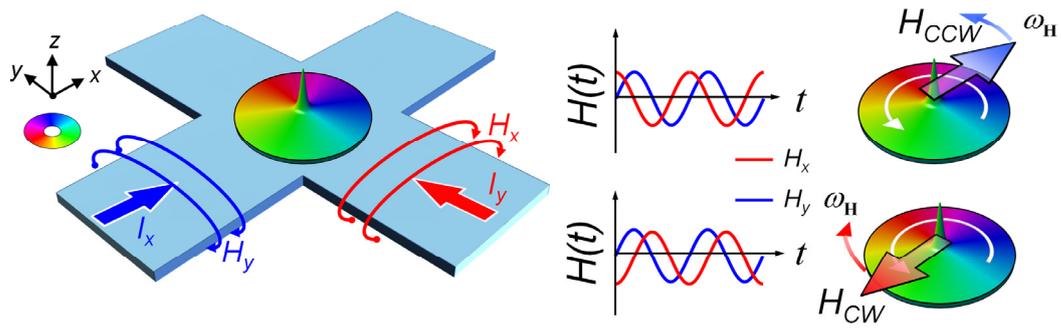

**FIG. 2.**

**(a)**

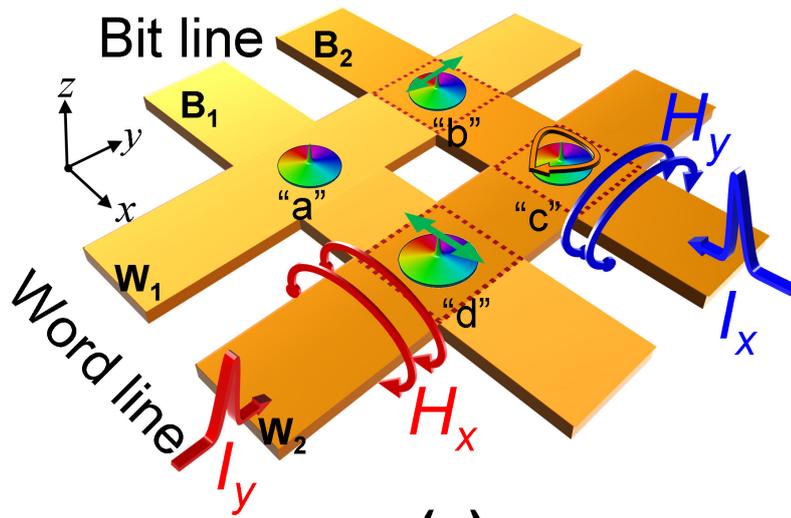

**(b)**

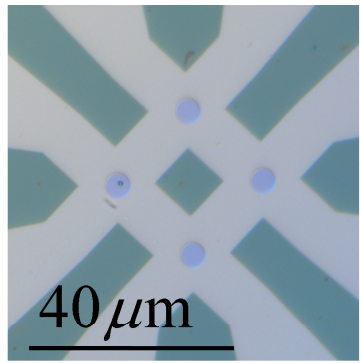

**(c)**

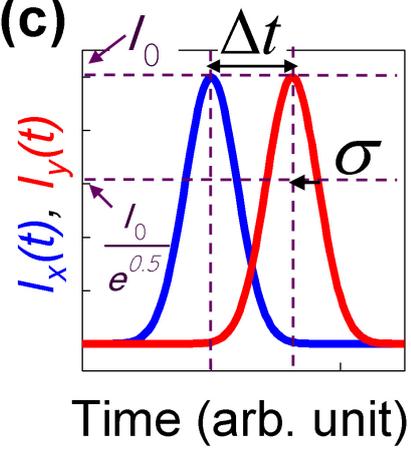

**FIG. 3.**

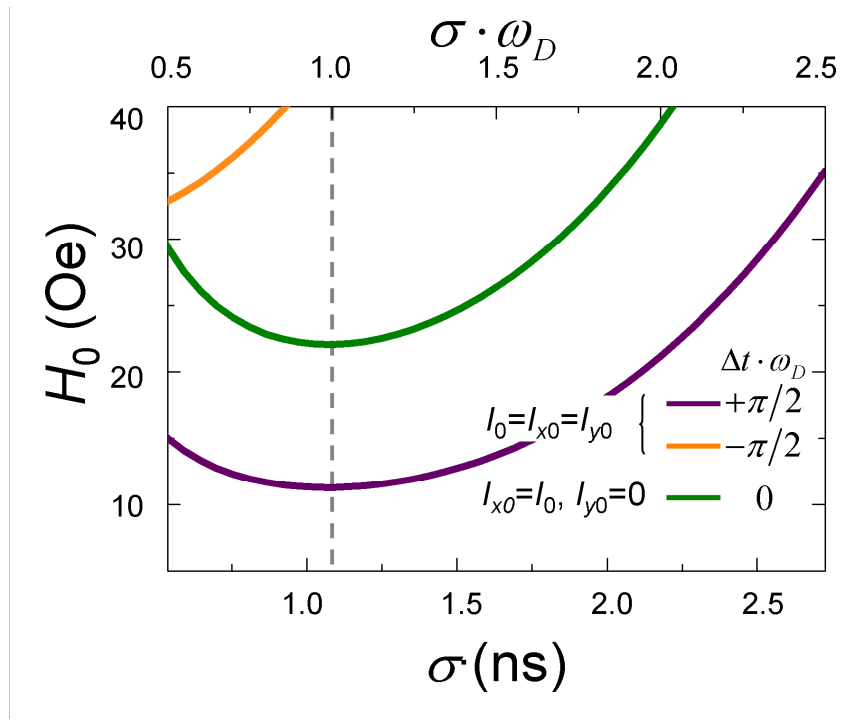

**FIG. 4.**

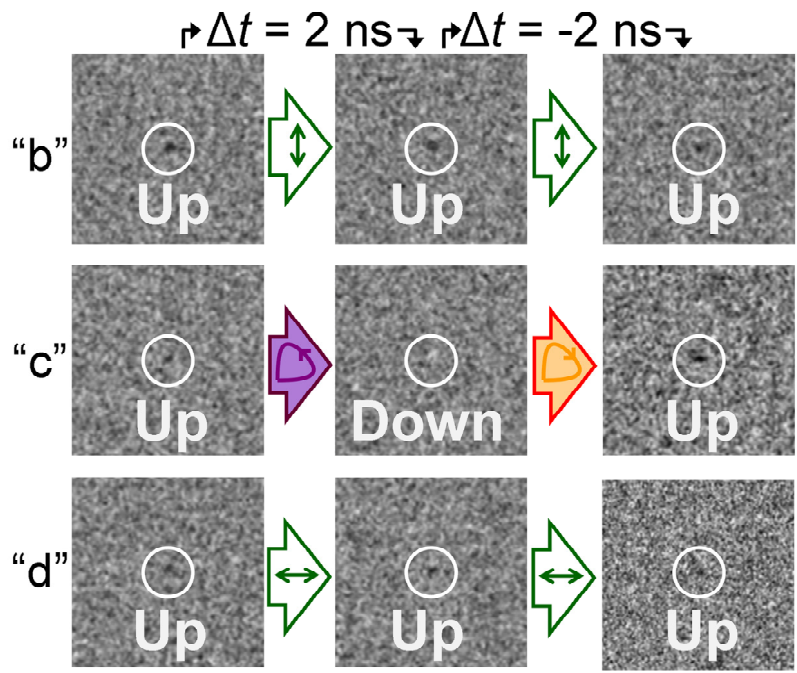